\documentclass[prd,aps,floatfix,superscriptaddress,footinbib,twocolumn,10pt,English]{revtex4}

\usepackage{amssymb,amsmath}
\usepackage{graphicx}
\usepackage{color}
\usepackage[colorlinks=true]{hyperref}
\usepackage{enumitem}
\usepackage{bm}

\usepackage{graphicx}
\graphicspath{ {./figures/} }

\newcommand{\grad}{\nabla}
\begin{document}  
	
	\title{Mimicking Kerr's multipole moments} 
	
		\author{B\'eatrice Bonga}
		\affiliation{Institute for Mathematics, Astrophysics and Particle Physics, Radboud University, 6525 AJ Nijmegen, The Netherlands}
		\author{Huan Yang}
		\affiliation{Perimeter Institute for Theoretical Physics, Waterloo, Ontario N2L 2Y5, Canada }
		\affiliation{University of Guelph, Guelph, Ontario N1G 2W1, Canada}

		\begin{abstract} 
			Multipole moments carry a lot of information about the gravitational field. Nonetheless, knowing all the multipole moments of an object does not determine conclusively the nature of the object itself. In particular, the field multipole moments of the Kerr spacetime are not unique. Here we construct several physically motivated Newtonian objects with multipole moments identical to those of Kerr. Moreover, we also provide a description of how to include post-Newtonian corrections to these objects without changing their multipole moments.
		\end{abstract}
		
		\maketitle 
		
		\section{ Introduction}		
		Ground-based gravitational-wave (GW) detectors have achieved tremendous success with the observation of merging stellar-mass black holes (BHs) and neutron stars (NSs). These observations are not only of astrophysical interest, but also interesting with regard to the foundations of gravity as they allow for stringent tests of general relativity. Such tests include searching for additional gravitational wave polarization modes \cite{Abbott:2017tlp,Callister:2017ocg,Isi:2017fbj,Chatziioannou:2021mij}, consistency of higher harmonics with the dominant harmonics in the signal \cite{Dhanpal:2018ufk,Kastha:2018bcr,Islam:2019dmk}, effects of dispersion during the wave propagation indicating a non-zero mass for the graviton \cite{Finn:2001qi,Mirshekari:2011yq,Perkins:2018tir}, searches for echoes \cite{Wang:2018gin,Testa:2018bzd,Micchi:2020gqy} and  parametrized tests \cite{Yunes:2009ke,Mishra:2010tp,Li:2011cg,Li:2011vx,Meidam:2017dgf,Carson:2020rea, LIGOScientific:2019fpa,testsGR-LIGO}. If the gravitational waves are emitted by a black hole, these tests aim to determine whether these black holes are described by the Kerr solutions of general relativity or some black hole solution in modified theories of gravity, or some other exotic compact objects altogether.
		
		%The ringdown spectrum is a superposition of quasi-normal modes of the merged remnant black hole. The oscillation frequency and decay time of each mode are completely determined by the mass $M$ and spin $a$ of the final Kerr black hole. Therefore, measuring as many modes as possible in the ringdown spectrum allows for a test of the nature of the final merged object and whether it truly is a black hole as described by the Kerr spacetime. Just as the quasi-normal modes of a Kerr black hole are completely determined by its mass and spin, so are all multipole moments. Therefore, the ringdown test is sometimes thought of as a test of the multipolar structure of the Kerr spacetime.

		In this spirit, a natural question to ask is: how unique is the exterior Kerr solution? To make this question more concrete, how unique are the multipole moments of the Kerr black hole? In other words, are there other (stellar) objects with the same multipole moments as those of the Kerr spacetime? 
		Some remarkable results in the literature answer a related question: if you know the multipole moments of a given spacetime, how much do you know about the spacetime itself? In the Newtonian theory, the equivalent question is trivially answered: knowing all multipole moments, one can directly reconstruct the gravitational potential outside sources given that the multipole moments are simply defined as the coefficients in the $1/r$ expansion of the gravitational potential. In the general relativistic context, the situation is  significantly more challenging, nonetheless similarly rigid results have been established. In particular, the Geroch-Hansen multipole moments characterize a stationary, vacuum spacetime uniquely up to isometries \cite{beig1980,Herberthson:2009ze}.\footnote{The restriction to stationary spacetimes is necessary for the Geroch-Hansen multipole moments to be defined.} Moreover, any stationary, asymptotically flat vacuum solution approaches the Kerr metric at infinity \cite{Beig:1980be}. 		
		These results are strong and suggest that any vacuum spacetime with \emph{all} field multipole moments equal to those of the Kerr spacetime has to be the Kerr spacetime itself. This seems to answer the question whether there are any objects with the same multipole moments of Kerr to the negative.

		However, these results all rely on a key assumption: the absence of sources. If we relax this condition, we show by an explicit construction that there are many stationary, axisymmetric Newtonian objects with identical multipole moments as those of a rotating black hole in general relativity. We also provide a constructive algorithm to go beyond the Newtonian context and include post-Newtonian corrections, but leave a fully relativistic generalization for future work (see \cite{McManus_1991,Bicak:1993zz,Pichon:1996pda,lynden-bell_2003,Will:2008ys} and references therein for earlier attempts at finding matter sources for the Kerr metric).\footnote{These earlier works are in the context of full non-linear general relativity and focus mostly on matching the Kerr metric, rather than just its multipolar structure at infinity. As a result, different approximations are made as well as different assumptions for the matter sources.}
		The objects constructed in this manner  satisfy the dominant energy condition $\rho \ge 0$ everywhere, although their stress  is not isotropic. In particular, material elements at different locations may not satisfy the same equation of state. While these objects may not exist in nature, this work shows explicitly that even if one knows all the (field) multipole moments of an object, one cannot conclusively tell the nature of the object.

		This also has theoretical implications on the conjecture put forward in \cite{Bianchi:2020bxa}, in which it is suggested that the multipole moments of a Kerr black hole are minimal in some sense. In particular, they provide numerical evidence that all multipole moments of a large family of horizonless microstate geometries known as fuzzballs are larger (in absolute value) than those of a Kerr black hole with the same mass and spin. The simple Newtonian objects we construct show that the Kerr values are not minimal, in the sense that Newtonian objects with the same mass and spin as a Kerr black hole  may have smaller multipole moments than the Kerr values. While their conjecture may still hold for the class of specific modified black hole solutions they considered, it cannot be more generically true for all possible compact objects within general relativity.\footnote{In fact, the minimalness conjecture rests on very specific families of microstate geometries \cite{counterexamples-microstate-geometries}. Generically, there is no such bound on the multipolar structure of microstate geometries. The counterexamples in \cite{counterexamples-microstate-geometries} are provided in the context of almost-BPS microstate geometries, which are slightly different from the supersymmetric ones in \cite{Bianchi:2020bxa}, but a construction of counterexamples in the SUSY setting is entirely analogous. Moreover, the results in  \cite{Bena:2020see,Bena:2020uup} also indirectly support the idea  that the minimalness conjecture does not hold for generic perturbations off the Kerr spacetime, at least within the framework of those papers, i.e. string theory black holes in 4 dimensions.}
		
		In recent years there is a revolutionary development of the Post-Newtonian and Post-Minkowskian theory based on effective field theory calculations of scattering amplitudes \cite{damour2020classical,bini2020sixth,bini2020binary}.
		In particular, it was shown that the scattering experiment of ``minimally'' coupled spin fields gives rise to the Kerr multipoles up to high orders in black hole spin \cite{vaidya2015gravitational,guevara2019holomorphic,cachazo2020leading,arkani2017scattering,guevara2019scattering,Aoude:2020mlg,Chung:2018kqs}. This is a rather surprising result as ``minimally coupled'' spin fields and Kerr black holes are rather different objects in nature. One possible way to build a connection between these two systems is to require that the Kerr moments are special in certain sense, e.g. minimal for any object with the same mass and angular momentum. The analysis presented in this work  provides counterexamples to such an intuitive explanation, as one can construct objects with smaller multipole moments than the Kerr values within general relativity. The coincidence observed in \cite{vaidya2015gravitational,guevara2019holomorphic,cachazo2020leading,arkani2017scattering,guevara2019scattering} ought to have a deeper origin.

		This paper is organized as follows. In Sec.~\ref{sec:many-multipole-moments}, we discuss the various notions of multipole moments and make sharp the comparison we make in this paper regarding the multipole moments of the Kerr spacetime and our Newtonian mimicker. Thorne's field multipole moments for the Kerr spacetime are also presented in that section. In Sec.~\ref{sec:newtonian}, we construct a Newtonian object whose field multipole moments have the same value as those of the Kerr exterior region. A generalization to include post-Newtonian corrections is described in Sec.~\ref{sec:post-Newtonian}. We conclude in Sec.~\ref{sec:discussion}.
		Our conventions are: We set Newton's constant $G$ and the speed of light $c$ both equal to one, the spacetime metric has signature -+++ and for the normalization of the various harmonics, we use the conventions in \cite{Thorne:1980ru}.
		
		%\vspace{0.2cm}	
		\section{Many multipole moments}
		\label{sec:many-multipole-moments}
		There are two distinct notions of multipole moments: source multipole moments, defined as integrals over the source, and field multipole moments, defined from the gravitational field near infinity. 		
		In Newtonian gravity, the source multipole moments describe the way in which mass is distributed, while the field multipole moments are important to determine the motion of extended or nearby objects. 	While the definitions are distinct, the source and field multipole moments of a single object are the same in the Newtonian context (see e.g. Eqs.~(1.139) and~(1.140) in \cite{pw}). 		
		In general relativity, the story is more complicated and generically one cannot even define the source multipole moments rigorously in the full non-linear context \cite{Dixon}. The exception are spacetimes with black holes described by axisymmetric isolated horizons (for which the symmetry restriction only applies to the horizon geometry and not the entire spacetime) \cite{Ashtekar:2004gp}. 
		Another exception is the scenario in which a post-Newtonian description of the sources applies \cite{Thorne:1980ru}.
		Field multipole moments are well-defined for a large class of spacetimes. The Geroch-Hansen multipole moments are an example of such field multipoles \cite{Geroch:1970cd,Hansen:1974zz}. 
		Here we will not use the geometric definition by Geroch and Hansen, but we use the definition due to Thorne instead \cite{Thorne:1980ru}. We make this choice because the Newtonian limit is more transparent in Thorne's approach. In addition, despite the clear difference in their definitions, Thorne's multipole moments of a stationary, asymptotically flat spacetime are identical to the field multipole moments of Geroch-Hansen, up to a normalization factor \cite{Gursel}. 
		
		The post-Newtonian field and source multipole moments are equivalent when no gravitational radiation is present. In this paper, we will use this equivalence to relate the (post-)Newtonian \emph{source} multipole moments of the objects we construct to the relativistic \emph{field} multipole moments of the Kerr spacetime.
				
		\subsection{Multipole moments of Kerr}
		The mass and current multipole moments of the Kerr spacetime are only non-zero for $m=0$ due to its axisymmetry and are given by:
		\begin{align}
			I^{\ell 0}_{\rm Kerr} \! \! &= \!
				\begin{cases}
					%\! M (i a)^\ell \sqrt{4\pi} \sqrt{\frac{(2\ell+1)(\ell+1)(\ell+2)}{2\ell (\ell-1)}} \frac{(2\ell)!}{2^{\ell -2}\ell!} \frac{1}{(2\ell+1)!! (2\ell-1)!!}\\
					\! M (i a)^\ell \sqrt{4\pi} \sqrt{\frac{(2\ell+1)(\ell+1)(\ell+2)}{2\ell (\ell-1)}} \frac{2^{\ell+1}(\ell-1)!}{(2\ell-1)!  (2\ell+1)}\\  %& \text{if $\ell$ is even} \\
					\! 0 %& \text{if $\ell$ is odd}
				\end{cases} \\
				S^{\ell 0}_{\rm Kerr}\! \! &= \!
				\begin{cases}
					\! 0  \\ %& \text{if $\ell$ is even} \\
					%\! i M (ia)^\ell \sqrt{4\pi} \sqrt{\frac{(2\ell+1)(\ell+1)(\ell+2)}{2\ell (\ell-1)}} \frac{(2\ell-1)!}{(\ell-1)! 2^{\ell-3}} \frac{1}{(2\ell+1)!! (2\ell-1)!!}
					\! i M (ia)^\ell \sqrt{4\pi} \sqrt{\frac{(2\ell+1)(\ell+1)(\ell+2)}{2\ell (\ell-1)}} \frac{2^{\ell+1}(\ell-1)!}{(2\ell-1)!  (2\ell+1)}
					%& \text{if $\ell$ is odd}
				\end{cases} 
		\end{align}
		where the first line indicates the result for $\ell$ being even and the second for $\ell$ being odd. Here, $M$ is the mass parameter of the Kerr spacetime while $\ell$ and $m$ indicate the degree and order of the spherical harmonic decomposition. The above formulas are only valid for $\ell \ge 2$. For $\ell=0$ the mass multipole moment of Kerr is simply $M$ and the current multipole is not defined, and for $\ell=1$ the mass multipole moment vanishes and the current multiple moment is $a M$. The vanishing of the odd mass multipole moments and even current multipole moments is due to the reflection-symmetry of Kerr in its equatorial plane. The dependence on $\ell$ has been derived using the relation between the Geroch-Hansen multipole moments and Thorne's symmetric trace-free multipole moments in \cite{Gursel} and translating those results to the multipole moments in the spherical harmonic decomposition. We have explicitly checked these numerical factors up to $\ell = 6$ using the ACMC-6 coordinate system in \cite{Sopuerta:2011te}. 
		
		%\vspace{0.2cm}		
		\section{A Newtonian mimicker} \label{sec:newtonian}
		
		In this section, we will first discuss the mass multipole moments and next consider the spin multipole moments. Since the Kerr spacetime is stationary, the configuration of the Newtonian object  should be time-independent, 
i.e., no explicit dependence on the time coordinate $t$.

		\subsection{Mass multipole moments}
		We start with a  general Newtonian star with a density profile $\rho(r,\theta,\phi)$. Decomposing the mass density as  
		\begin{equation}
			\rho(r, \theta, \phi)= \sum_{\ell, m} \rho_{\ell m}(r) Y_{\ell m}(\theta,\phi), \label{eq:mass-density}
		\end{equation}
		the Newtonian mass multipoles are given by
		\begin{equation}
		I^{\ell m}_{\rm Newtonian} = \frac{16 \pi}{(2 \ell + 1)!} \sqrt{\frac{(\ell+2)(\ell+1)}{2\ell (\ell -1)}}	 \int \! dr \; r^{\ell +2 } \rho_{\ell m}(r) \; .	 \label{eq:mass-multipoles}
		\end{equation}

\begin{figure}
			\begin{flushleft}
				\includegraphics[width=0.45\textwidth]{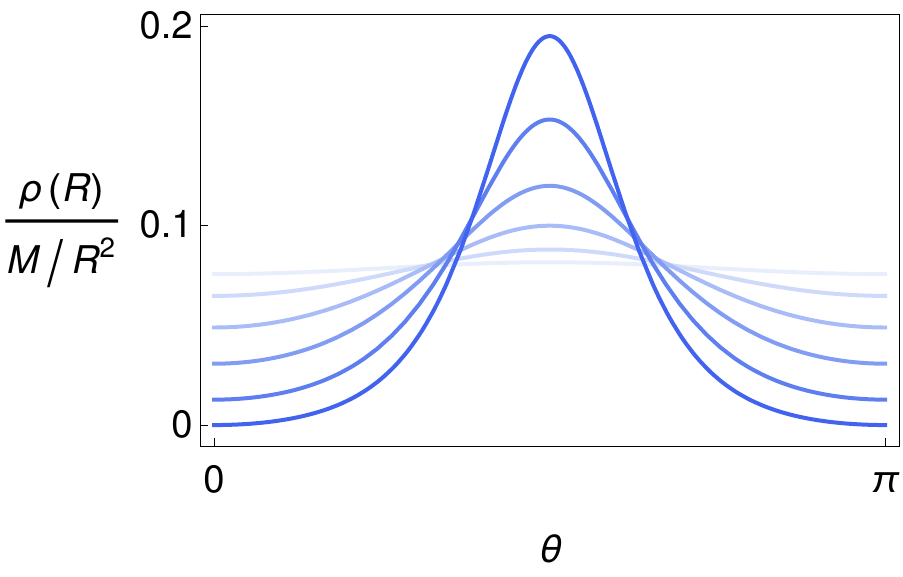} 
			\end{flushleft}
			\includegraphics[width=0.4\textwidth]{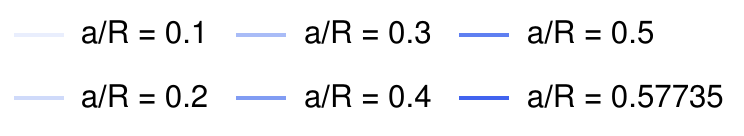}\caption{This figure shows the angular dependence of the mass density $\rho$ for the thin shell model at radius $R$ for different ratios of $a/R$. It is clear that the mass density is positive for all $a/R \le 0.57735$. In all cases, the mass density is peaked near the equator and minimal at the poles. For plotting purposes, the sum over $\ell$ in Eq.~\eqref{eq:mass-density} is truncated at $\ell=200$. This truncation does not alter the plot as higher order terms are suppressed (in fact, for $a/R=0.1$ truncating at $\ell = 4$ does not alter the plot, but as $a/R$ increases more terms are needed).}
			\label{fig:rhoversustheta}
		\end{figure}

		One immediate observation of Eq.~\eqref{eq:mass-multipoles}  is that there are many possible $\rho_{\ell m}$ giving rise to the same set of $I^{\ell m}_{\rm Newtonian}$, because of the nature of the integration equation. In this study we shall focus on the axis-symmetric scenarios, with no explicit dependence on $\phi$. As a first example, we assume a thin shell of matter with mass density $\rho_{\ell m} = \alpha_{\ell} \delta_{m,0} \frac{M}{R^2} \left(\frac{a}{R}\right)^\ell \delta(r-R)$ with $R$ being the radius of the thin shell, $M$ its mass and $a$ its spin parameter. The constant $\alpha_\ell$ is determined by matching the Newtonian mass multipole moments with those of the Kerr black hole 
		\begin{equation}
			\alpha_\ell = \begin{cases}
				%\sqrt{\frac{2\ell+1}{4\pi}} \frac{(2\ell)!}{2^\ell \ell!} \frac{i^\ell}{(2\ell-1)!!} =
				i^\ell \sqrt{\frac{2\ell+1}{4\pi}} \\
				0
			\end{cases} .
		\end{equation} 
		For this choice, whenever $\rho_{\ell m}(R)$ is not zero, it is positive in roughly half of the cases ($\ell=0,4,8,12,\ldots$) and negative in the other half ($\ell=2,6,10,14,\ldots$). Similarly, the combination $\rho_{\ell m}(r) Y_{\ell m}(\theta,\phi)$ for given $\ell$ and $m=0$ can be negative at specific polar angles even if $\rho_{\ell m}$ is positive. With the monopole piece ($\ell=0$) included, the mass density itself $\rho(r,\theta,\phi)$ is  nowhere negative as long as $a/R \le 0.57735$. This is illustrated in Fig.~\ref{fig:rhoversustheta}, which shows that the mass density is concentrated near the equator and minimized near its pole. As the spin to radius ratio increases, this feature becomes more pronounced. The $\ell$ multipoles of the mass density of the thin shell at the poles (where $\phi$ is undefined) is given by
		\begin{align}
			\rho_{\ell 0}(r=R) Y_{\ell 0}(\theta=0,\phi) &= i^\ell \frac{2\ell+1}{4\pi} \left(\frac{a}{R}\right)^\ell \; .
		\end{align}
		When the ratio $a/R$ is increased beyond the numerical value 0.57735, the mass density becomes negative near the poles.
		The restriction on the ratio $a/R$ certainly seems reasonable to assume, because if we were to take the spin parameter to be equal to a maximally spinning Kerr black hole (i.e., $a=M$), the compactness of this object as measured by the ratio $M/R$ would be close to a half. In this regime, one certainly would need relativistic corrections as this is close to the compactness of a Schwarzschild black hole. 
		
		%The restriction on the ratio $a/R$ certainly seems reasonable to assume, because if we take $R$ to be very small so that the ratio $a/R$ is large, i.e., we take $R$ as compact as a Kerr black hole so that $R = M + M \sqrt{1 - \chi^2}$ with $\chi= a/M$, then 
	%	\begin{equation}
	%		\frac{a}{R} = \frac{\chi}{1+ \sqrt{1-\chi^2}}
	%	\end{equation}
	%	which is less than $0.57735$ if $\chi \le 0.866025$. If $R$ were to be as compact as a Kerr black hole, we would 
		
\begin{figure}			
			\begin{flushleft}
				\includegraphics[width=0.45\textwidth]{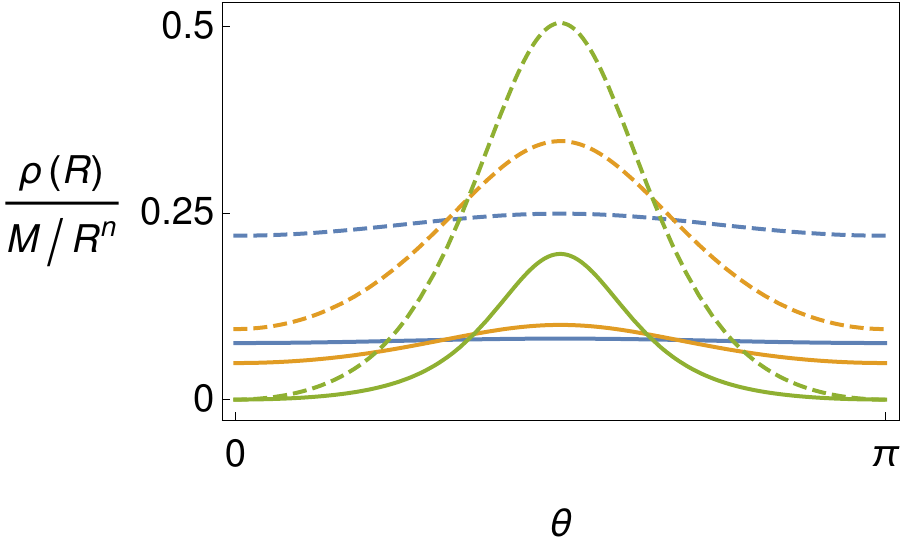} 
			\end{flushleft}
			\includegraphics[width=0.48\textwidth]{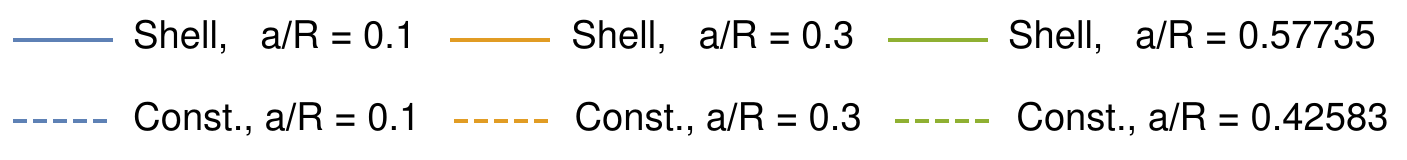} 
			\caption{This figure shows the angular dependence of the mass density $\rho$ evaluated at $r=R$ for the thin shell model versus the constant density profile for different ratios of $a/R$. The value of $n$ on the vertical axis is 2 for the thin shell model and 3 for the constant density profile.}
			\label{fig:rhocomparison}
		\end{figure}

		Other straight-forward examples with mass multiple moments equal to those of the Kerr spacetime include a stellar object with a constant radial profile and an elementary decaying profile:
		\begin{align*}
			\rho_{\ell m} = \begin{cases}
				\beta_\ell \delta_{m,0} \frac{M}{R^3} \left(\frac{a}{R}\right)^\ell & \text{constant radial profile} \\
				\gamma_\ell \delta_{m,0} \frac{M}{R^3} \left(\frac{a}{R}\right)^\ell \cos \left(\frac{\pi r}{4 R} \right)& \text{decaying radial profile} 
			\end{cases}
		\end{align*}
		with the constant $\beta_\ell$ and $\gamma_\ell$ given by
		\begin{align*}
			\beta_\ell & = \left(\ell+3\right) \alpha_\ell \\
			\gamma_\ell & = \frac{\pi^{3+\ell}}{64 \cdot 2^{2\ell}} \frac{1}{\text{Re}\left[i^{\ell+1} \Gamma(\ell+3,- \frac{i \pi}{4})\right]+ \Gamma(3+\ell) \sin \left(\frac{\pi \ell}{2}\right)} \alpha_{\ell} \; .
		\end{align*}
		For simplicity, we will only consider the thin shell model and the constant radial profile from hereon. The mass density for the constant radial density profile is non-negative for all polar angles as long as $a/R \le 0.42583$. A comparison with the thin shell model is shown in Fig.~\ref{fig:rhocomparison}.

		\subsection{Current multipole moments}
		As a second step, we match the values of the current multipole moments. For a generic Newtonian body, they are given by
		\begin{align}
			S^{\ell m}_{\rm Newtonian} &= \frac{32 \pi}{(2 \ell +1)!!}\sqrt{\frac{(\ell+2)(2\ell+1)}{2 (\ell+1)(\ell-1)}} \times \notag \\
			&\quad \int d^2 \Omega \int d r \; r^{\ell+2 } \; \epsilon_{ijk} \bar{Y}^i_{\ell-1,\ell m} v^j \rho \, n^k  \label{eq:current-multipoles}
		\end{align}
		where $n^i$ is the radial unit vector and $\bar{Y}^i_{\ell-1,\ell m} = (-1)^m Y^i_{\ell-1,\ell -m}$ are the pure-orbital harmonics. The latter are related to the spin-weighted spherical harmonics $_{-1}Y_{\ell -m}$ through
		\begin{equation}
			Y^i_{\ell-1,\ell-m} = \sqrt{\frac{\ell+1}{2(2\ell+1)}} \left(_{-1}Y_{\ell -m} m^i - _{1}Y_{\ell-m} \bar{m}^i\right)
		\end{equation} 
		with $m^i$ and its complex conjugate $\bar{m}^i$ being the Newman-Penrose complex null vectors on the sphere (for more details, see \cite[Sec.~2]{Thorne:1980ru}). 
		The velocity field is most intuitively decomposed into the pure-spin vector harmonics:
		\begin{equation}
			v^i = \sum_{\ell, m} \left[R_{\ell m} {}^{R}Y_{\ell m}^i 
			+ E_{\ell m} {}^{E}Y_{\ell m}^i  + B_{\ell m} {}^{B}Y_{\ell m}^i  \right]
		\end{equation}
		where ${}^{R}Y_{\ell m}^i, {}^{E}Y_{\ell m}^i$ and ${}^{B}Y_{\ell m}^i$ are normalized as in \cite[Eq.~(2.18)]{Thorne:1980ru}. Due to the cross product of $v^i$ with the radial unit vector $n^i$ in the definition of the current multipole moments, $R_{\ell m}$ does not contribute to $S^{\ell m}_{\rm Newtonian}$ and thus is not constrained by the requirement that the current multipole moments are equal to those of Kerr. The $E_{\ell m}$ contribute an imaginary part to the current multipole moments when $m$ is odd, and are thus required to vanish in order to match the current multipole moments of Kerr: $E_{\ell m} =0$. Therefore, the only relevant coefficients are $B_{\ell m}$. 
		Nevertheless, the coefficients $R_{\ell m}$ are not entirely free as the continuity equation constrains its behavior. For both the thin shell model and the constant density profile, the continuity equation implies that the divergence of the velocity vector field has to vanish. Since the part proportional to $B_{\ell m}$ is by construction divergence-free, this imposes the following constraint on $R_{\ell m}$
		\begin{equation}
				\frac{d}{dr} R_{\ell m} + \frac{2}{r} R_{\ell m} = 0 \; .
		\end{equation} 
		A solution to this equation is $R_{\ell m}(r) \sim 1/r^2$, but this is not well-defined at the origin and therefore we are required to set $R_{\ell m} =0$ for the model with a constant radial profile. We will also set $R_{\ell m}=0$ for the thin shell model.
		
		The $B_{\ell m}$ part in the velocity field can also be written in terms of the spin-weighted harmonics as:
		\begin{equation}
			v^i_B = - \frac{i}{\sqrt{2}} \sum_{\ell, m} B_{\ell m}(r) \left[ {}_{-1}Y_{\ell m}(\theta,\phi) m^i + {}_{1}Y_{\ell m}(\theta,\phi) \bar{m}^i \right] \; . 
			\label{eq:vBi}
		\end{equation}
		Substituting this decomposition as well as the decomposition for the mass density in Eq.~\eqref{eq:mass-density} into the current multipole moments in Eq.~\eqref{eq:current-multipoles}, and using that $\epsilon_{ijk}n^k = 2 i \; m_{[i} \bar{m}_{j]}$, we obtain
		\begin{align}
				S^{\ell m}_{\rm Newtonian} &= \frac{32 \pi}{(2 \ell +1)!!}\sqrt{\frac{(\ell+2)(2\ell+1)}{2 (\ell+1)(\ell-1)}} \;  \times \notag \\
			&  \int d r \; r^{\ell+2 } \frac{(-1)^m}{2} \sqrt{\frac{\ell+1}{2\ell+1}} \sum_{\ell', m'
			} \sum_{\ell'',m''} \rho_{\ell'm'} B_{\ell''m''} \notag \\
			&  \int d^2 \Omega \; \left( 	
			{}_{0}Y_{\ell'm'} \; {}_{-1}Y_{\ell''m''} \; {}_{1}Y_{\ell-m} \right. \notag \\
			&\left. \qquad \qquad + \; 
			{}_{0}Y_{\ell'm'} \; {}_{1}Y_{\ell''m''} \; {}_{-1}Y_{\ell-m} 
				 \right) \; .
		\end{align}
		The angular integrals over the two sets of three spin-weighted spherical harmonics are given by the product of two $3j$-symbols\cite[Eq.~(34.3.22)]{NIST:DLMF}. Since $\rho_{\ell m}$ is only non-zero for $m=0$ and $\ell$ even, the $3j$-symbols simplify significantly. Moreover, as we assume these current multipole moments to match those of a rotating black hole,  this expression can be further simplified by setting $B_{\ell m}$ equal to zero for all $m \neq 0$ and $\ell$ even. After these simplifications, the infinite sums over $\ell'$ and $\ell''$ still remain:
		\begin{align}
			S^{\ell m}_{\rm Newtonian}&= \frac{32 \pi}{(2 \ell +1)!!}\sqrt{\frac{(\ell+2)(2\ell+1)}{2 (\ell+1)(\ell-1)}} \times \notag \\
			&\quad  \int d r \; r^{\ell+2 } \frac{(-1)^m}{2} \sqrt{\frac{\ell+1}{2\ell+1}} \sum_{\ell',\ell''
			}  \rho_{\ell'0} B_{\ell''0} \notag \\
			& \quad \sqrt{\frac{(2\ell+1)(2\ell'+1)(2\ell''+1)}{4\pi}} 
			\begin{pmatrix}
				\ell' & \ell'' & \ell \\
				0 & 0 & 0
			\end{pmatrix} \notag \\
			& \quad \left[ 
			\begin{pmatrix}
				\ell' & \ell'' & \ell \\
				0 & 1 & -1
			\end{pmatrix} + 
			\begin{pmatrix}
				\ell' & \ell'' & \ell \\
				0 & -1 & 1
			\end{pmatrix} 
			\right].
		\end{align}
		Setting these current multipole moments equal to those of Kerr yields a very large but invertible matrix equation for $B_{\ell m}$, which is in principle solvable. Here we will only provide a perturbative solution in $a/R$ up to $\mathcal{O} \left(\frac{a}{R}\right)^6$. For both the thin-shell model and the model with a constant radial profile, we find
		\begin{align}
			\label{eq:solB}
				B_{\ell 0} &= \begin{cases}
					\sqrt{6\pi} \frac{a}{R} \left[1- \left(\frac{a}{R}\right)^2 + \frac{11}{7} \left(\frac{a}{R}\right)^4 \right] + \mathcal{O} \left(\frac{a}{R}\right)^6 % & \ell = 1 
					\\
					\sqrt{\frac{16\pi}{21}} \left(\frac{a}{R}\right)^3 \left[1+ \left(\frac{a}{R}\right)^2\right]+ \mathcal{O} \left(\frac{a}{R}\right)^6 % & \ell = 3 
					\\
					\frac{16}{7} \sqrt{\frac{2\pi}{165}} \left(\frac{a}{R}\right)^5 + \mathcal{O} \left(\frac{a}{R}\right)^6 %& \ell=5 
					\\
					\ldots
				\end{cases}
		\end{align}
		with the different lines indicating the results for $\ell=1,3,5,\ldots$. (Both models have the same velocity vector field as the extra factor of $\ell +3$ in $\beta_{\ell}$ compared to $\alpha_{\ell}$ is canceled after performing the integration over $r$.)
		Using Mathematica, these expressions are easily obtained for much larger $\ell$ (we do not show these here as the expressions are not particularly informative). It is clear from this particular solution that the coefficients $B_{\ell 0}$ are all of the form
		\begin{equation}
			B_{\ell 0} = \sum_{\ell'=\ell}^{\infty} c_{\ell'}^{(\ell)} \left(\frac{a}{R}\right)^{\ell'}
		\end{equation}
		with $c_{\ell'}^{\ell}$ numerical coefficients that are non-zero only for $\ell, \ell'$ odd. For instance, we already know from Eq.~\eqref{eq:solB} that $c_{1}^{(1)} = \sqrt{6\pi}, c_{3}^{(1)} = - \sqrt{6\pi}, c_{5}^{(1)} = \frac{11}{7}\sqrt{6\pi} $ and $c_{1}^{(3)} = \sqrt{16\pi/21}$.
				
%		An alternative perturbative calculation to obtain the coefficients $B_{\ell m}$ is to only account for the leading order term in the mass density $\rho_{00}$ first, solve for $B_{\ell 0}$ in closed form, and then take higher order terms in $\rho_{\ell m}$ into account to find corrections to $B_{\ell 0}$. This approach yields after the first step:
%		\begin{align*}
%			B_{\ell 0}^{\rm approx} &= - i \left(\frac{ia}{R}\right)^\ell \sqrt{4\pi} \sqrt{\frac{(2\ell+1)(\ell+1)}{\ell}}\; \times \\ &\qquad \frac{(2\ell -1)!}{(2\ell-1)!!} \frac{1}{(\ell-1)! 2^\ell} 
%		\end{align*}
%		Note that this expression only agrees with the the leading order term in Eq.~\eqref{eq:solB} for $\ell=1$. This is as expected, as for instance for $B_{50}$ terms with $\rho_{00} S_{50}, \rho_{20} S_{30}, \rho_{20}^2 S_{10}$ and $\rho_{40} S_{10}$ all  contribute to the leading order term in $(a/R)^5$.
%		\textcolor{blue}{Decided to remove this as I could not obtain closed form expressions at the second iteration?}

\begin{figure}
				\begin{flushleft}
				\includegraphics[width=0.45\textwidth]{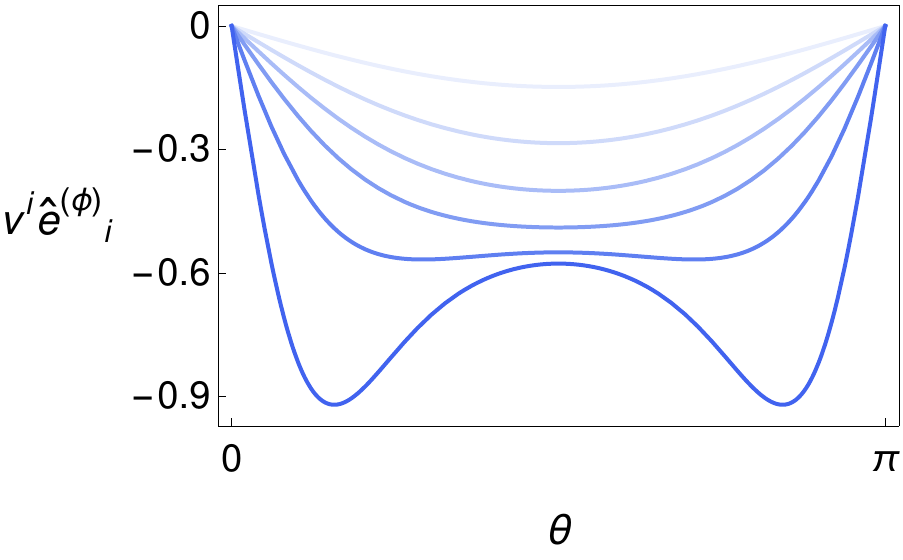} 
			\end{flushleft}
			\includegraphics[width=0.4\textwidth]{legend2-rhoversustheta.pdf}\caption{This plot shows the angular dependence of $\phi$-component of the velocity $v_B^i$ in an orthonormal basis for different ratios of $a/R$. The maximal velocity increases as the ratio $a/R$ increases, but is always less than the speed of light. To make this plot, we truncated the sum over $\ell$ in Eq.~\eqref{eq:vBi} at $\ell =20$ and verified that the higher order terms do not change the result. (For the model with a constant radial profile, the ratio $a/R$ needs to be restricted to $a/R \le 0.42583$.)}
			\label{fig:plotvphi}
		\end{figure}

		Substituting the solution for $B_{\ell 0}$ into the expression for the velocity field, we find that the only non-zero component of the velocity vector field is given by
		\begin{align}
		v^i_B \hat{e}^{(\phi)}_i & =
			\frac{1}{2} \left[ -3 \frac{a}{R}+ \left(\frac{a}{R}\right)^3 \left(4-5 \cos ^2\theta \right)  \right.  \\
			& \quad \left.   -\left(\frac{a}{R}\right)^5 \left(6 \cos ^4\theta +\cos^2\theta +4\right)  \right] \sin \theta + \mathcal{O} \left(\frac{a}{R}\right)^6 \notag 
		\end{align}
		where $\hat{e}_i^{(\phi)}$ is the orthonormal vector in the $\phi$-direction.
		Fig.~\ref{fig:plotvphi} shows the angular dependence of the velocity above for different values of $a/R$ (for plotting purposes, we used expressions accurate to $\mathcal{O}\left(\frac{a}{R}\right)^{20}$). This also shows that the magnitude of the velocity $v^i$ is less than the speed of light for all polar angles, and it vanishes at the poles.

		%\vspace{0.2cm}		
		\section{Going the extra mile: including post-Newtonian corrections}
		\label{sec:post-Newtonian}
		In the previous sections, we have demonstrated how to construct Newtonian objects with identical moments to those of Kerr. The construction is not unique and there are in fact many possible objects which share the same values for the multipole moments as those of Kerr.
		In this section, we  give a constructive argument how to extend those results to include post-Newtonian corrections. 
		
		First note that the expression for the mass multipole moments with leading order post-Newtonian corrections involves significantly more terms than at Newtonian order \cite[Eq.~(5.31)]{Thorne:1980ru}
		\begin{align}
			I^{\ell m}& = \frac{16 \pi}{(2 \ell + 1)!} \sqrt{\frac{(\ell+2)(\ell+1)}{2\ell (\ell -1)}}	 \int \! d^2 \Omega \int \! dr \; r^{\ell +2 } \notag \\
			&\quad \left[\tau_{00} \bar{Y}^{\ell m} + \frac{1}{2} \sqrt{\frac{(2\ell+1)\ell(\ell-1)}{2\ell -1}} \bar{T}_{ij}^{2 \; \; \ell-2, \ell m} \tau^{ij} \right. \notag \\
			&\quad \left.  \sqrt{\frac{6\ell (\ell-1)^2}{(2\ell+3)(\ell+1)(2\ell -1)}}\bar{T}_{ij}^{2 \; \; \ell, \ell m} \tau^{ij}  \right] 
		\end{align}
		where $\bar{T}_{ij}^{2 \; \; \ell(-2), \ell m}$ are tensor harmonics \cite[Eq.~(2.27)-(2.28)]{Thorne:1980ru} and $\tau_{\mu \nu}$ is the effective stress-energy tensor evaluated at the first post-Newtonian order in the post-Newtonian de Donger gauge so that
		\begin{subequations}
			\begin{align}	
			\tau_{00} & = \rho \left(1 + 4 U + v^2 + \Pi \right) - \frac{3}{8\pi} \grad_k U \grad^k U \\
			\tau_{ij} & = \rho\; v_i v_j - \frac{1}{4\pi} \grad_i U \grad_j U - \frac{1}{2\pi} U \grad_i \grad_j U \notag \\
			& \quad + \left(P + \frac{3}{8\pi} \grad_k U \grad^k U - 2 \rho U \right) g_{ij} - 2 \Sigma_{ij} \; .
			\end{align}
		\end{subequations}
		Here the mass density $\rho$ is the Newtonian mass density and $v^i$ the Newtonian velocity vector field. The Newtonian potential $U$ and pressure $P$ as well as the specific internal energy density $\Pi$ and stress tensor $\Sigma_{ij}$ are determined by the Poisson, Euler and conservation of energy equation
		\begin{align}
			&\grad^2 U  = - 4 \pi \; \rho  \label{eq:poisson}\\
			&\grad_i P  - 2 \grad_j \Sigma_{ij} = \rho \grad_i U + \rho  \, \Omega^2 \left(n_i - r \cos \theta \; \hat{e}^{(z)}_i \right) \label{eq:euler}\\
			& \rho \; v^i \grad_i \Pi = - P \grad_i v^i \; , \label{eq:cons-energy}
		\end{align} 
		where $\hat{e}^{(z)}_i$ is the unit-vector in the $z$-direction.  The second term on the right hand side of the Euler equation is the centrifugal force for which the angular velocity  $\Omega$ is determined by $\Omega^2 = v_B^i v^B_i/r^2$ (and thus, $\Omega$ also depends on the polar angle $\theta$).	In the above equations, we restricted ourselves to the case in which all fields are time-independent.		
		
		Since the leading order terms themselves already match the multipole moments of Kerr, the post-Newtonian corrections have to vanish. Therefore, we need to establish whether there is enough functional freedom to ensure this. The strategy is to take $\rho$ as the energy density previously obtained plus some perturbation at post-Newtonian level, say $\delta \rho$, chosen such that exactly all the additional terms introduced at the first post-Newtonian order vanish. Concretely, one first needs to solve for $U$ given $\rho$ in Eq.~\eqref{eq:poisson} (a solution is easily constructed by decomposing $U$ into spherical harmonics and using \cite[Eq.~(1.128)]{pw})
		\begin{align}
			U_{\ell m} = \frac{4\pi}{2\ell+1} &\left[ r^\ell \int_r^\infty \rho_{\ell m}(r') r'^{-\ell+1} dr' \right. \notag \\
			&\left. \quad  + \frac{1}{r^{\ell+1}} \int_0^r  \rho_{\ell m}(r')r'^{\ell+2} dr' \right] \; .
		\end{align}
		This Newtonian potential will then serve as a source for the pressure $P$ and stress tensor $\Sigma_{ij}$ in Eq.~\eqref{eq:euler}. Since the stress tensor is symmetric and traceless by definition and we are interested in the axially symmetric case, we can decompose the stress tensor simply as a linear combination of four tensor harmonics  
		\begin{align}
			\Sigma_{ij} &= \sum_{\ell} \left[ 
			\sigma^{(1)}_{\ell m} \; Y_{\ell m} \; \left(g_{ij} - 3 n_i n_j\right) \right. \notag \\
			&\left. \quad + \sigma_{\ell m}^{(2)} \left(_{-1}Y_{\ell m} \left(m_i n_j + n_i m_j \right)- \, {}_{1}Y_{\ell m} \left(\bar{m}_i n_j + n_i \bar{m}_j \right) \right) \right. \notag \\
			&\left. \quad + 
			\sigma_{\ell m}^{(3)} \left(_{-2}Y_{\ell m} m_i m_j + \, {}_{2}Y_{\ell m} \bar{m}_i \bar{m}_j \right) \right] 
		\end{align}
		with $\sigma^{(i)}_{\ell m} = 0$ whenever $m \neq 0$. In the case of the thin shell model, the stress tensor satisfies $\Sigma_{ij} n^j = 0$ and $\sigma_{lm}^{(1)} = 0$.
		The $\phi$-component of the Euler equation is trivially satisfied, but the $\theta$- and $r$-component yield two differential equations. Since the source on the right-hand side of these equations is the product of $\rho$ and $\grad_i U$, we obtain $3j$-symbols again which result into large matrix equations (similar to the situation for $B_{\ell 0}$). These equations can also be solved to any desired order using, for instance, Mathematica.
		The solutions are not unique given that there are two equations and four free functions $P_{\ell m}$ and $\sigma_{\ell m}^{(i)}$ (three in the case of the thin shell model). One can use this freedom to simplify the solutions.
		A simple solution for $\Pi$ in Eq.~\eqref{eq:cons-energy} would be to take $\Pi=0$ (as $v^i$ is divergence free). 
		Knowing all the relevant components of $\tau_{\mu \nu}$, one can finally calculate the leading order post-Newtonian correction to the mass multipole moments. The last step is to determine $\delta \rho$ such that
		\begin{align}
			\frac{16 \pi}{(2 \ell + 1)!} \sqrt{\frac{(\ell+2)(\ell+1)}{2\ell (\ell -1)}}	 \int \! dr \; r^{\ell +2 } \delta\rho_{\ell m}(r) = - I^{\ell m}_{\rm 1 PN} \; .
		\end{align}
		The same argument applies to the current multipole moments.
		This algorithm shows that the Newtonian construction in Sec.~\ref{sec:newtonian} can be easily extended to the leading post-Newtonian order. The expectation is that this will also hold for higher-order post-Newtonian corrections.

		%\vspace{0.2cm}		
		\section{Discussion}
		\label{sec:discussion}		
		The field multipole moments of the Kerr spacetime are not unique: we constructed several examples in the Newtonian theory with identical multipole moments as those of a Kerr black hole. 
		Therefore, knowing all the (field) multipole moments of an object does not conclusively tell us the nature of the object. That we were able to construct such examples is not surprising in light of the fact that the uniqueness results mentioned in the Introduction rests on ellipticity of the field equations. The differential equation for the Newtonian potential is clearly elliptic. Einstein's equations are elliptic provided that they describe stationary vacuum spacetimes and are formulated on the manifold of trajectories of the time-like Killing vector field (and written in suitable coordinates)\cite{Friedrich:2006km,hagen_1970}.\footnote{The ellipticity also applies to the conformally completed spacetime constructed for the formulation of the Geroch-Hansen multipole moments.} Accordingly, uniqueness typically fails when ellipticity of the equations is lost. The presence of matter is one such way in which ellipticity and consequently uniqueness no longer applies.

		The construction of the explicit examples in this paper also shows that the Kerr multipole moments are not minimal in the sense that their absolute value is minimized for objects with the same mass and angular momentum, as the construction in this paper can also be used to find objects with multipole moments smaller than those of the Kerr spacetime.  
		This point can be further elaborated by considering an example of a solid star (so that non-isotropic stress is allowed) with a mountain in the north pole. If the star is rotating, the spin-induced quadrupole may perfectly cancel the mountain-associated quadrupole, so that there is no net quadrupole moment for such star. It is straightforward to see that such construction leads to a quadrupole moment smaller than the Kerr value, assuming the mass and angular momentum are the same.
		Of course, the ``minimalness'' conjecture may still hold if we restrict the matter sources to be fluid stars.

		There are several important drawbacks to this analysis. First, to mimic the multipole moments of very compact objects with $R$ not much greater than $M$, the Newtonian analysis is not enough and one is required to include post-Newtonian corrections (possibly many).   This complicates the analysis, but is in principle calculable. 
		
		Second, the star profiles that mimic Kerr moments suggest materials with non-isotropic stress and likely position-dependent equation of state, or different types of materials at different locations. This ``naturalness" problem may be used to argue against the likelihood of finding such object(s) in nature, as the required compositions are difficult to be naturally fabricated. It is however worth noting that the naturalness problem can be split into two sub-problems. The first one is whether the object is allowed by the laws of nature (e.g., general relativity), and the second one is  whether its formation is natural, without the intervention of intelligence. Our study can only address the first question. Despite this possible objection, it is still of fundamental interest whether one can also construct fully relativistic objects with the same multipole moment structure as those of the Kerr spacetime. 
		
		Third, while the objects we constructed have a non-negative mass density everywhere and velocity smaller than the speed of light, we did not investigate their stability under generic linear perturbations. 	Such a task should be more technically complicated than analyzing the perturbation of barotropic stars because of the non-isotropic stress and non-homogeneous equation of state.	
		We shall leave this for future work.		 
		
	%	An interesting question is whether one can also construct fully relativistic objects with the same multipole moment structure as those of the Kerr spacetime. We leave this for future work.

		\section*{Acknowledgments} 	
		We thank Eric Poisson for reading a draft of this paper. We also would like to thank him and Vojt\v{e}ch Witzany for pointing us to \cite{McManus_1991,Bicak:1993zz,Pichon:1996pda,lynden-bell_2003,Will:2008ys} and the general program of finding matter sources for the Kerr solution. B.B. also thanks Daniel Mayerson for enlightening discussions on microstate geometries and the minimalness conjecture. H. Y. is supported by the Natural Sciences and Engineering Research Council of Canada and in part by Perimeter Institute for Theoretical Physics. Research at Perimeter Institute is supported in part by the Government of Canada through the Department of Innovation, Science and Economic Development Canada and by the Province of Ontario through the Ministry of Colleges and Universities.

		%%%%%%%%%%%%%%%%%%%%%%%%%%%%%%%%%%%%%%%%%%%%
		\bibliography{ref-mimicking-kerr-v2}

\begin{thebibliography}{54}
\expandafter\ifx\csname natexlab\endcsname\relax\def\natexlab#1{#1}\fi
\expandafter\ifx\csname bibnamefont\endcsname\relax
  \def\bibnamefont#1{#1}\fi
\expandafter\ifx\csname bibfnamefont\endcsname\relax
  \def\bibfnamefont#1{#1}\fi
\expandafter\ifx\csname citenamefont\endcsname\relax
  \def\citenamefont#1{#1}\fi
\expandafter\ifx\csname url\endcsname\relax
  \def\url#1{\texttt{#1}}\fi
\expandafter\ifx\csname urlprefix\endcsname\relax\def\urlprefix{URL }\fi
\providecommand{\bibinfo}[2]{#2}
\providecommand{\eprint}[2][]{\url{#2}}

\bibitem[{\citenamefont{Abbott et~al.}(2018)}]{Abbott:2017tlp}
\bibinfo{author}{\bibfnamefont{B.~P.} \bibnamefont{Abbott}}
  \bibnamefont{et~al.} (\bibinfo{collaboration}{LIGO Scientific, Virgo}),
  \bibinfo{journal}{Phys. Rev. Lett.} \textbf{\bibinfo{volume}{120}},
  \bibinfo{pages}{031104} (\bibinfo{year}{2018}), \eprint{1709.09203}.

\bibitem[{\citenamefont{Callister et~al.}(2017)\citenamefont{Callister,
  Biscoveanu, Christensen, Isi, Matas, Minazzoli, Regimbau, Sakellariadou,
  Tasson, and Thrane}}]{Callister:2017ocg}
\bibinfo{author}{\bibfnamefont{T.}~\bibnamefont{Callister}},
  \bibinfo{author}{\bibfnamefont{A.~S.} \bibnamefont{Biscoveanu}},
  \bibinfo{author}{\bibfnamefont{N.}~\bibnamefont{Christensen}},
  \bibinfo{author}{\bibfnamefont{M.}~\bibnamefont{Isi}},
  \bibinfo{author}{\bibfnamefont{A.}~\bibnamefont{Matas}},
  \bibinfo{author}{\bibfnamefont{O.}~\bibnamefont{Minazzoli}},
  \bibinfo{author}{\bibfnamefont{T.}~\bibnamefont{Regimbau}},
  \bibinfo{author}{\bibfnamefont{M.}~\bibnamefont{Sakellariadou}},
  \bibinfo{author}{\bibfnamefont{J.}~\bibnamefont{Tasson}}, \bibnamefont{and}
  \bibinfo{author}{\bibfnamefont{E.}~\bibnamefont{Thrane}},
  \bibinfo{journal}{Phys. Rev. X} \textbf{\bibinfo{volume}{7}},
  \bibinfo{pages}{041058} (\bibinfo{year}{2017}), \eprint{1704.08373}.

\bibitem[{\citenamefont{Isi and Weinstein}(2017)}]{Isi:2017fbj}
\bibinfo{author}{\bibfnamefont{M.}~\bibnamefont{Isi}} \bibnamefont{and}
  \bibinfo{author}{\bibfnamefont{A.~J.} \bibnamefont{Weinstein}}
  (\bibinfo{year}{2017}), \eprint{1710.03794}.

\bibitem[{\citenamefont{Chatziioannou et~al.}(2021)\citenamefont{Chatziioannou,
  Isi, Haster, and Littenberg}}]{Chatziioannou:2021mij}
\bibinfo{author}{\bibfnamefont{K.}~\bibnamefont{Chatziioannou}},
  \bibinfo{author}{\bibfnamefont{M.}~\bibnamefont{Isi}},
  \bibinfo{author}{\bibfnamefont{C.-J.} \bibnamefont{Haster}},
  \bibnamefont{and} \bibinfo{author}{\bibfnamefont{T.~B.}
  \bibnamefont{Littenberg}} (\bibinfo{year}{2021}), \eprint{2105.01521}.

\bibitem[{\citenamefont{Dhanpal et~al.}(2019)\citenamefont{Dhanpal, Ghosh,
  Mehta, Ajith, and Sathyaprakash}}]{Dhanpal:2018ufk}
\bibinfo{author}{\bibfnamefont{S.}~\bibnamefont{Dhanpal}},
  \bibinfo{author}{\bibfnamefont{A.}~\bibnamefont{Ghosh}},
  \bibinfo{author}{\bibfnamefont{A.~K.} \bibnamefont{Mehta}},
  \bibinfo{author}{\bibfnamefont{P.}~\bibnamefont{Ajith}}, \bibnamefont{and}
  \bibinfo{author}{\bibfnamefont{B.~S.} \bibnamefont{Sathyaprakash}},
  \bibinfo{journal}{Phys. Rev. D} \textbf{\bibinfo{volume}{99}},
  \bibinfo{pages}{104056} (\bibinfo{year}{2019}), \eprint{1804.03297}.

\bibitem[{\citenamefont{Kastha et~al.}(2018)\citenamefont{Kastha, Gupta, Arun,
  Sathyaprakash, and Van Den~Broeck}}]{Kastha:2018bcr}
\bibinfo{author}{\bibfnamefont{S.}~\bibnamefont{Kastha}},
  \bibinfo{author}{\bibfnamefont{A.}~\bibnamefont{Gupta}},
  \bibinfo{author}{\bibfnamefont{K.~G.} \bibnamefont{Arun}},
  \bibinfo{author}{\bibfnamefont{B.~S.} \bibnamefont{Sathyaprakash}},
  \bibnamefont{and} \bibinfo{author}{\bibfnamefont{C.}~\bibnamefont{Van
  Den~Broeck}}, \bibinfo{journal}{Phys. Rev. D} \textbf{\bibinfo{volume}{98}},
  \bibinfo{pages}{124033} (\bibinfo{year}{2018}), \eprint{1809.10465}.

\bibitem[{\citenamefont{Islam et~al.}(2020)\citenamefont{Islam, Mehta, Ghosh,
  Varma, Ajith, and Sathyaprakash}}]{Islam:2019dmk}
\bibinfo{author}{\bibfnamefont{T.}~\bibnamefont{Islam}},
  \bibinfo{author}{\bibfnamefont{A.~K.} \bibnamefont{Mehta}},
  \bibinfo{author}{\bibfnamefont{A.}~\bibnamefont{Ghosh}},
  \bibinfo{author}{\bibfnamefont{V.}~\bibnamefont{Varma}},
  \bibinfo{author}{\bibfnamefont{P.}~\bibnamefont{Ajith}}, \bibnamefont{and}
  \bibinfo{author}{\bibfnamefont{B.~S.} \bibnamefont{Sathyaprakash}},
  \bibinfo{journal}{Phys. Rev. D} \textbf{\bibinfo{volume}{101}},
  \bibinfo{pages}{024032} (\bibinfo{year}{2020}), \eprint{1910.14259}.

\bibitem[{\citenamefont{Finn and Sutton}(2002)}]{Finn:2001qi}
\bibinfo{author}{\bibfnamefont{L.~S.} \bibnamefont{Finn}} \bibnamefont{and}
  \bibinfo{author}{\bibfnamefont{P.~J.} \bibnamefont{Sutton}},
  \bibinfo{journal}{Phys. Rev. D} \textbf{\bibinfo{volume}{65}},
  \bibinfo{pages}{044022} (\bibinfo{year}{2002}), \eprint{gr-qc/0109049}.

\bibitem[{\citenamefont{Mirshekari et~al.}(2012)\citenamefont{Mirshekari,
  Yunes, and Will}}]{Mirshekari:2011yq}
\bibinfo{author}{\bibfnamefont{S.}~\bibnamefont{Mirshekari}},
  \bibinfo{author}{\bibfnamefont{N.}~\bibnamefont{Yunes}}, \bibnamefont{and}
  \bibinfo{author}{\bibfnamefont{C.~M.} \bibnamefont{Will}},
  \bibinfo{journal}{Phys. Rev. D} \textbf{\bibinfo{volume}{85}},
  \bibinfo{pages}{024041} (\bibinfo{year}{2012}), \eprint{1110.2720}.

\bibitem[{\citenamefont{Perkins and Yunes}(2019)}]{Perkins:2018tir}
\bibinfo{author}{\bibfnamefont{S.}~\bibnamefont{Perkins}} \bibnamefont{and}
  \bibinfo{author}{\bibfnamefont{N.}~\bibnamefont{Yunes}},
  \bibinfo{journal}{Class. Quant. Grav.} \textbf{\bibinfo{volume}{36}},
  \bibinfo{pages}{055013} (\bibinfo{year}{2019}), \eprint{1811.02533}.

\bibitem[{\citenamefont{Wang and Afshordi}(2018)}]{Wang:2018gin}
\bibinfo{author}{\bibfnamefont{Q.}~\bibnamefont{Wang}} \bibnamefont{and}
  \bibinfo{author}{\bibfnamefont{N.}~\bibnamefont{Afshordi}},
  \bibinfo{journal}{Phys. Rev. D} \textbf{\bibinfo{volume}{97}},
  \bibinfo{pages}{124044} (\bibinfo{year}{2018}), \eprint{1803.02845}.

\bibitem[{\citenamefont{Testa and Pani}(2018)}]{Testa:2018bzd}
\bibinfo{author}{\bibfnamefont{A.}~\bibnamefont{Testa}} \bibnamefont{and}
  \bibinfo{author}{\bibfnamefont{P.}~\bibnamefont{Pani}},
  \bibinfo{journal}{Phys. Rev. D} \textbf{\bibinfo{volume}{98}},
  \bibinfo{pages}{044018} (\bibinfo{year}{2018}), \eprint{1806.04253}.

\bibitem[{\citenamefont{Longo~Micchi et~al.}(2021)\citenamefont{Longo~Micchi,
  Afshordi, and Chirenti}}]{Micchi:2020gqy}
\bibinfo{author}{\bibfnamefont{L.~F.} \bibnamefont{Longo~Micchi}},
  \bibinfo{author}{\bibfnamefont{N.}~\bibnamefont{Afshordi}}, \bibnamefont{and}
  \bibinfo{author}{\bibfnamefont{C.}~\bibnamefont{Chirenti}},
  \bibinfo{journal}{Phys. Rev. D} \textbf{\bibinfo{volume}{103}},
  \bibinfo{pages}{044028} (\bibinfo{year}{2021}), \eprint{2010.14578}.

\bibitem[{\citenamefont{Yunes and Pretorius}(2009)}]{Yunes:2009ke}
\bibinfo{author}{\bibfnamefont{N.}~\bibnamefont{Yunes}} \bibnamefont{and}
  \bibinfo{author}{\bibfnamefont{F.}~\bibnamefont{Pretorius}},
  \bibinfo{journal}{Phys. Rev. D} \textbf{\bibinfo{volume}{80}},
  \bibinfo{pages}{122003} (\bibinfo{year}{2009}), \eprint{0909.3328}.

\bibitem[{\citenamefont{Mishra et~al.}(2010)\citenamefont{Mishra, Arun, Iyer,
  and Sathyaprakash}}]{Mishra:2010tp}
\bibinfo{author}{\bibfnamefont{C.~K.} \bibnamefont{Mishra}},
  \bibinfo{author}{\bibfnamefont{K.~G.} \bibnamefont{Arun}},
  \bibinfo{author}{\bibfnamefont{B.~R.} \bibnamefont{Iyer}}, \bibnamefont{and}
  \bibinfo{author}{\bibfnamefont{B.~S.} \bibnamefont{Sathyaprakash}},
  \bibinfo{journal}{Phys. Rev. D} \textbf{\bibinfo{volume}{82}},
  \bibinfo{pages}{064010} (\bibinfo{year}{2010}), \eprint{1005.0304}.

\bibitem[{\citenamefont{Li et~al.}(2012{\natexlab{a}})\citenamefont{Li,
  Del~Pozzo, Vitale, Van Den~Broeck, Agathos, Veitch, Grover, Sidery, Sturani,
  and Vecchio}}]{Li:2011cg}
\bibinfo{author}{\bibfnamefont{T.~G.~F.} \bibnamefont{Li}},
  \bibinfo{author}{\bibfnamefont{W.}~\bibnamefont{Del~Pozzo}},
  \bibinfo{author}{\bibfnamefont{S.}~\bibnamefont{Vitale}},
  \bibinfo{author}{\bibfnamefont{C.}~\bibnamefont{Van Den~Broeck}},
  \bibinfo{author}{\bibfnamefont{M.}~\bibnamefont{Agathos}},
  \bibinfo{author}{\bibfnamefont{J.}~\bibnamefont{Veitch}},
  \bibinfo{author}{\bibfnamefont{K.}~\bibnamefont{Grover}},
  \bibinfo{author}{\bibfnamefont{T.}~\bibnamefont{Sidery}},
  \bibinfo{author}{\bibfnamefont{R.}~\bibnamefont{Sturani}}, \bibnamefont{and}
  \bibinfo{author}{\bibfnamefont{A.}~\bibnamefont{Vecchio}},
  \bibinfo{journal}{Phys. Rev. D} \textbf{\bibinfo{volume}{85}},
  \bibinfo{pages}{082003} (\bibinfo{year}{2012}{\natexlab{a}}),
  \eprint{1110.0530}.

\bibitem[{\citenamefont{Li et~al.}(2012{\natexlab{b}})\citenamefont{Li,
  Del~Pozzo, Vitale, Van Den~Broeck, Agathos, Veitch, Grover, Sidery, Sturani,
  and Vecchio}}]{Li:2011vx}
\bibinfo{author}{\bibfnamefont{T.~G.~F.} \bibnamefont{Li}},
  \bibinfo{author}{\bibfnamefont{W.}~\bibnamefont{Del~Pozzo}},
  \bibinfo{author}{\bibfnamefont{S.}~\bibnamefont{Vitale}},
  \bibinfo{author}{\bibfnamefont{C.}~\bibnamefont{Van Den~Broeck}},
  \bibinfo{author}{\bibfnamefont{M.}~\bibnamefont{Agathos}},
  \bibinfo{author}{\bibfnamefont{J.}~\bibnamefont{Veitch}},
  \bibinfo{author}{\bibfnamefont{K.}~\bibnamefont{Grover}},
  \bibinfo{author}{\bibfnamefont{T.}~\bibnamefont{Sidery}},
  \bibinfo{author}{\bibfnamefont{R.}~\bibnamefont{Sturani}}, \bibnamefont{and}
  \bibinfo{author}{\bibfnamefont{A.}~\bibnamefont{Vecchio}},
  \bibinfo{journal}{J. Phys. Conf. Ser.} \textbf{\bibinfo{volume}{363}},
  \bibinfo{pages}{012028} (\bibinfo{year}{2012}{\natexlab{b}}),
  \eprint{1111.5274}.

\bibitem[{\citenamefont{Meidam et~al.}(2018)}]{Meidam:2017dgf}
\bibinfo{author}{\bibfnamefont{J.}~\bibnamefont{Meidam}} \bibnamefont{et~al.},
  \bibinfo{journal}{Phys. Rev. D} \textbf{\bibinfo{volume}{97}},
  \bibinfo{pages}{044033} (\bibinfo{year}{2018}), \eprint{1712.08772}.

\bibitem[{\citenamefont{Carson and Yagi}(2020)}]{Carson:2020rea}
\bibinfo{author}{\bibfnamefont{Z.}~\bibnamefont{Carson}} \bibnamefont{and}
  \bibinfo{author}{\bibfnamefont{K.}~\bibnamefont{Yagi}}
  (\bibinfo{year}{2020}), \eprint{2011.02938}.

\bibitem[{\citenamefont{Abbott et~al.}(2019)}]{LIGOScientific:2019fpa}
\bibinfo{author}{\bibfnamefont{B.~P.} \bibnamefont{Abbott}}
  \bibnamefont{et~al.} (\bibinfo{collaboration}{LIGO Scientific, Virgo}),
  \bibinfo{journal}{Phys. Rev. D} \textbf{\bibinfo{volume}{100}},
  \bibinfo{pages}{104036} (\bibinfo{year}{2019}), \eprint{1903.04467}.

\bibitem[{\citenamefont{Abbott et~al.}(2020)}]{testsGR-LIGO}
\bibinfo{author}{\bibfnamefont{R.}~\bibnamefont{Abbott}} \bibnamefont{et~al.}
  (\bibinfo{collaboration}{LIGO Scientific, Virgo}) (\bibinfo{year}{2020}),
  \eprint{2010.14529}.

\bibitem[{\citenamefont{Beig and Simon}(1980{\natexlab{a}})}]{beig1980}
\bibinfo{author}{\bibfnamefont{R.}~\bibnamefont{Beig}} \bibnamefont{and}
  \bibinfo{author}{\bibfnamefont{W.}~\bibnamefont{Simon}},
  \bibinfo{journal}{Comm. Math. Phys.} \textbf{\bibinfo{volume}{78}},
  \bibinfo{pages}{75} (\bibinfo{year}{1980}{\natexlab{a}}),
  \urlprefix\url{https://projecteuclid.org:443/euclid.cmp/1103908502}.

\bibitem[{\citenamefont{Herberthson}(2009)}]{Herberthson:2009ze}
\bibinfo{author}{\bibfnamefont{M.}~\bibnamefont{Herberthson}},
  \bibinfo{journal}{Class. Quant. Grav.} \textbf{\bibinfo{volume}{26}},
  \bibinfo{pages}{215009} (\bibinfo{year}{2009}), \eprint{0906.4247}.

\bibitem[{\citenamefont{Beig and Simon}(1980{\natexlab{b}})}]{Beig:1980be}
\bibinfo{author}{\bibfnamefont{R.}~\bibnamefont{Beig}} \bibnamefont{and}
  \bibinfo{author}{\bibfnamefont{W.}~\bibnamefont{Simon}},
  \bibinfo{journal}{Gen. Rel. Grav.} \textbf{\bibinfo{volume}{12}},
  \bibinfo{pages}{1003} (\bibinfo{year}{1980}{\natexlab{b}}).

\bibitem[{\citenamefont{McManus}(1991)}]{McManus_1991}
\bibinfo{author}{\bibfnamefont{D.}~\bibnamefont{McManus}},
  \bibinfo{journal}{Classical and Quantum Gravity}
  \textbf{\bibinfo{volume}{8}}, \bibinfo{pages}{863} (\bibinfo{year}{1991}),
  \urlprefix\url{https://doi.org/10.1088/0264-9381/8/5/011}.

\bibitem[{\citenamefont{Bicak and Ledvinka}(1993)}]{Bicak:1993zz}
\bibinfo{author}{\bibfnamefont{J.}~\bibnamefont{Bicak}} \bibnamefont{and}
  \bibinfo{author}{\bibfnamefont{T.}~\bibnamefont{Ledvinka}},
  \bibinfo{journal}{Phys. Rev. Lett.} \textbf{\bibinfo{volume}{71}},
  \bibinfo{pages}{1669} (\bibinfo{year}{1993}).

\bibitem[{\citenamefont{Pichon and Lynden-Bell}(1996)}]{Pichon:1996pda}
\bibinfo{author}{\bibfnamefont{C.}~\bibnamefont{Pichon}} \bibnamefont{and}
  \bibinfo{author}{\bibfnamefont{D.}~\bibnamefont{Lynden-Bell}},
  \bibinfo{journal}{Mon. Not. Roy. Astron. Soc.}
  \textbf{\bibinfo{volume}{280}}, \bibinfo{pages}{1007} (\bibinfo{year}{1996}),
  \eprint{astro-ph/9605037}.

\bibitem[{\citenamefont{Lynden-Bell}(2003)}]{lynden-bell_2003}
\bibinfo{author}{\bibfnamefont{D.}~\bibnamefont{Lynden-Bell}},
  \emph{\bibinfo{title}{A magic electromagnetic field}}
  (\bibinfo{publisher}{Cambridge University Press}, \bibinfo{year}{2003}), p.
  \bibinfo{pages}{369–376}.

\bibitem[{\citenamefont{Will}(2009)}]{Will:2008ys}
\bibinfo{author}{\bibfnamefont{C.~M.} \bibnamefont{Will}},
  \bibinfo{journal}{Phys. Rev. Lett.} \textbf{\bibinfo{volume}{102}},
  \bibinfo{pages}{061101} (\bibinfo{year}{2009}), \eprint{0812.0110}.

\bibitem[{\citenamefont{Bianchi et~al.}(2020)\citenamefont{Bianchi, Consoli,
  Grillo, Morales, Pani, and Raposo}}]{Bianchi:2020bxa}
\bibinfo{author}{\bibfnamefont{M.}~\bibnamefont{Bianchi}},
  \bibinfo{author}{\bibfnamefont{D.}~\bibnamefont{Consoli}},
  \bibinfo{author}{\bibfnamefont{A.}~\bibnamefont{Grillo}},
  \bibinfo{author}{\bibfnamefont{J.~F.} \bibnamefont{Morales}},
  \bibinfo{author}{\bibfnamefont{P.}~\bibnamefont{Pani}}, \bibnamefont{and}
  \bibinfo{author}{\bibfnamefont{G.}~\bibnamefont{Raposo}},
  \bibinfo{journal}{Phys. Rev. Lett.} \textbf{\bibinfo{volume}{125}},
  \bibinfo{pages}{221601} (\bibinfo{year}{2020}), \eprint{2007.01743}.

\bibitem[{\citenamefont{Damour}(2020)}]{damour2020classical}
\bibinfo{author}{\bibfnamefont{T.}~\bibnamefont{Damour}},
  \bibinfo{journal}{Physical Review D} \textbf{\bibinfo{volume}{102}},
  \bibinfo{pages}{024060} (\bibinfo{year}{2020}).

\bibitem[{\citenamefont{Bini et~al.}(2020{\natexlab{a}})\citenamefont{Bini,
  Damour, and Geralico}}]{bini2020sixth}
\bibinfo{author}{\bibfnamefont{D.}~\bibnamefont{Bini}},
  \bibinfo{author}{\bibfnamefont{T.}~\bibnamefont{Damour}}, \bibnamefont{and}
  \bibinfo{author}{\bibfnamefont{A.}~\bibnamefont{Geralico}},
  \bibinfo{journal}{Physical Review D} \textbf{\bibinfo{volume}{102}},
  \bibinfo{pages}{024061} (\bibinfo{year}{2020}{\natexlab{a}}).

\bibitem[{\citenamefont{Bini et~al.}(2020{\natexlab{b}})\citenamefont{Bini,
  Damour, and Geralico}}]{bini2020binary}
\bibinfo{author}{\bibfnamefont{D.}~\bibnamefont{Bini}},
  \bibinfo{author}{\bibfnamefont{T.}~\bibnamefont{Damour}}, \bibnamefont{and}
  \bibinfo{author}{\bibfnamefont{A.}~\bibnamefont{Geralico}},
  \bibinfo{journal}{Physical Review D} \textbf{\bibinfo{volume}{102}},
  \bibinfo{pages}{024062} (\bibinfo{year}{2020}{\natexlab{b}}).

\bibitem[{\citenamefont{Vaidya}(2015)}]{vaidya2015gravitational}
\bibinfo{author}{\bibfnamefont{V.}~\bibnamefont{Vaidya}},
  \bibinfo{journal}{Physical Review D} \textbf{\bibinfo{volume}{91}},
  \bibinfo{pages}{024017} (\bibinfo{year}{2015}).

\bibitem[{\citenamefont{Guevara}(2019)}]{guevara2019holomorphic}
\bibinfo{author}{\bibfnamefont{A.}~\bibnamefont{Guevara}},
  \bibinfo{journal}{Journal of High Energy Physics}
  \textbf{\bibinfo{volume}{2019}}, \bibinfo{pages}{33} (\bibinfo{year}{2019}).

\bibitem[{\citenamefont{Cachazo and Guevara}(2020)}]{cachazo2020leading}
\bibinfo{author}{\bibfnamefont{F.}~\bibnamefont{Cachazo}} \bibnamefont{and}
  \bibinfo{author}{\bibfnamefont{A.}~\bibnamefont{Guevara}},
  \bibinfo{journal}{Journal of High Energy Physics}
  \textbf{\bibinfo{volume}{2020}}, \bibinfo{pages}{1} (\bibinfo{year}{2020}).

\bibitem[{\citenamefont{Arkani-Hamed et~al.}(2017)\citenamefont{Arkani-Hamed,
  Huang, and Huang}}]{arkani2017scattering}
\bibinfo{author}{\bibfnamefont{N.}~\bibnamefont{Arkani-Hamed}},
  \bibinfo{author}{\bibfnamefont{T.-C.} \bibnamefont{Huang}}, \bibnamefont{and}
  \bibinfo{author}{\bibfnamefont{Y.-t.} \bibnamefont{Huang}},
  \bibinfo{journal}{arXiv preprint arXiv:1709.04891}  (\bibinfo{year}{2017}).

\bibitem[{\citenamefont{Guevara et~al.}(2019)\citenamefont{Guevara, Ochirov,
  and Vines}}]{guevara2019scattering}
\bibinfo{author}{\bibfnamefont{A.}~\bibnamefont{Guevara}},
  \bibinfo{author}{\bibfnamefont{A.}~\bibnamefont{Ochirov}}, \bibnamefont{and}
  \bibinfo{author}{\bibfnamefont{J.}~\bibnamefont{Vines}},
  \bibinfo{journal}{Journal of High Energy Physics}
  \textbf{\bibinfo{volume}{2019}}, \bibinfo{pages}{1} (\bibinfo{year}{2019}).

\bibitem[{\citenamefont{Aoude et~al.}(2020)\citenamefont{Aoude, Chung, Huang,
  Machado, and Tam}}]{Aoude:2020mlg}
\bibinfo{author}{\bibfnamefont{R.}~\bibnamefont{Aoude}},
  \bibinfo{author}{\bibfnamefont{M.-Z.} \bibnamefont{Chung}},
  \bibinfo{author}{\bibfnamefont{Y.-t.} \bibnamefont{Huang}},
  \bibinfo{author}{\bibfnamefont{C.~S.} \bibnamefont{Machado}},
  \bibnamefont{and} \bibinfo{author}{\bibfnamefont{M.-K.} \bibnamefont{Tam}},
  \bibinfo{journal}{Phys. Rev. Lett.} \textbf{\bibinfo{volume}{125}},
  \bibinfo{pages}{181602} (\bibinfo{year}{2020}), \eprint{2007.09486}.

\bibitem[{\citenamefont{Chung et~al.}(2019)\citenamefont{Chung, Huang, Kim, and
  Lee}}]{Chung:2018kqs}
\bibinfo{author}{\bibfnamefont{M.-Z.} \bibnamefont{Chung}},
  \bibinfo{author}{\bibfnamefont{Y.-T.} \bibnamefont{Huang}},
  \bibinfo{author}{\bibfnamefont{J.-W.} \bibnamefont{Kim}}, \bibnamefont{and}
  \bibinfo{author}{\bibfnamefont{S.}~\bibnamefont{Lee}},
  \bibinfo{journal}{JHEP} \textbf{\bibinfo{volume}{04}}, \bibinfo{pages}{156}
  (\bibinfo{year}{2019}), \eprint{1812.08752}.

\bibitem[{\citenamefont{Thorne}(1980)}]{Thorne:1980ru}
\bibinfo{author}{\bibfnamefont{K.~S.} \bibnamefont{Thorne}},
  \bibinfo{journal}{Rev. Mod. Phys.} \textbf{\bibinfo{volume}{52}},
  \bibinfo{pages}{299} (\bibinfo{year}{1980}).

\bibitem[{\citenamefont{Poisson and Will}(2014)}]{pw}
\bibinfo{author}{\bibfnamefont{E.}~\bibnamefont{Poisson}} \bibnamefont{and}
  \bibinfo{author}{\bibfnamefont{C.}~\bibnamefont{Will}},
  \emph{\bibinfo{title}{{Gravity, Newtonian, Post-Newtonian, Relativistic}}}
  (\bibinfo{publisher}{Cambridge University Press}, \bibinfo{year}{2014}).

\bibitem[{\citenamefont{Dixon}(1974)}]{Dixon}
\bibinfo{author}{\bibfnamefont{W.~G.} \bibnamefont{Dixon}},
  \bibinfo{journal}{Philosophical Transactions of the Royal Society of London.
  Series A, Mathematical and Physical Sciences} \textbf{\bibinfo{volume}{277}},
  \bibinfo{pages}{59} (\bibinfo{year}{1974}).

\bibitem[{\citenamefont{Ashtekar et~al.}(2004)\citenamefont{Ashtekar, Engle,
  Pawlowski, and Van Den~Broeck}}]{Ashtekar:2004gp}
\bibinfo{author}{\bibfnamefont{A.}~\bibnamefont{Ashtekar}},
  \bibinfo{author}{\bibfnamefont{J.}~\bibnamefont{Engle}},
  \bibinfo{author}{\bibfnamefont{T.}~\bibnamefont{Pawlowski}},
  \bibnamefont{and} \bibinfo{author}{\bibfnamefont{C.}~\bibnamefont{Van
  Den~Broeck}}, \bibinfo{journal}{Class. Quant. Grav.}
  \textbf{\bibinfo{volume}{21}}, \bibinfo{pages}{2549} (\bibinfo{year}{2004}),
  \eprint{gr-qc/0401114}.

\bibitem[{\citenamefont{Geroch}(1970)}]{Geroch:1970cd}
\bibinfo{author}{\bibfnamefont{R.~P.} \bibnamefont{Geroch}},
  \bibinfo{journal}{J. Math. Phys.} \textbf{\bibinfo{volume}{11}},
  \bibinfo{pages}{2580} (\bibinfo{year}{1970}).

\bibitem[{\citenamefont{Hansen}(1974)}]{Hansen:1974zz}
\bibinfo{author}{\bibfnamefont{R.~O.} \bibnamefont{Hansen}},
  \bibinfo{journal}{J. Math. Phys.} \textbf{\bibinfo{volume}{15}},
  \bibinfo{pages}{46} (\bibinfo{year}{1974}).

\bibitem[{\citenamefont{{G{\"u}rsel}}(1983)}]{Gursel}
\bibinfo{author}{\bibfnamefont{Y.}~\bibnamefont{{G{\"u}rsel}}},
  \bibinfo{journal}{General Relativity and Gravitation}
  \textbf{\bibinfo{volume}{15}}, \bibinfo{pages}{737} (\bibinfo{year}{1983}).

\bibitem[{\citenamefont{Sopuerta and Yunes}(2011)}]{Sopuerta:2011te}
\bibinfo{author}{\bibfnamefont{C.~F.} \bibnamefont{Sopuerta}} \bibnamefont{and}
  \bibinfo{author}{\bibfnamefont{N.}~\bibnamefont{Yunes}},
  \bibinfo{journal}{Phys. Rev. D} \textbf{\bibinfo{volume}{84}},
  \bibinfo{pages}{124060} (\bibinfo{year}{2011}), \eprint{1109.0572}.

\bibitem[{{\relax DLMF}()}]{NIST:DLMF}
{\relax DLMF}, \emph{\bibinfo{title}{{\it NIST Digital Library of Mathematical
  Functions}}}, \bibinfo{howpublished}{http://dlmf.nist.gov/, Release 1.1.1 of
  2021-03-15}, \bibinfo{note}{f.~W.~J. Olver, A.~B. {Olde Daalhuis}, D.~W.
  Lozier, B.~I. Schneider, R.~F. Boisvert, C.~W. Clark, B.~R. Miller, B.~V.
  Saunders, H.~S. Cohl, and M.~A. McClain, eds.},
  \urlprefix\url{http://dlmf.nist.gov/}.

\bibitem[{\citenamefont{Friedrich}(2007)}]{Friedrich:2006km}
\bibinfo{author}{\bibfnamefont{H.}~\bibnamefont{Friedrich}},
  \bibinfo{journal}{Annales Henri Poincare} \textbf{\bibinfo{volume}{8}},
  \bibinfo{pages}{817} (\bibinfo{year}{2007}), \eprint{gr-qc/0606133}.

\bibitem[{\citenamefont{Hagen}(1970)}]{hagen_1970}
\bibinfo{author}{\bibfnamefont{H.~M.~z.} \bibnamefont{Hagen}},
  \bibinfo{journal}{Mathematical Proceedings of the Cambridge Philosophical
  Society} \textbf{\bibinfo{volume}{68}}, \bibinfo{pages}{199–201}
  (\bibinfo{year}{1970}).

\bibitem[{\citenamefont{Bah et~al.}(2021)\citenamefont{Bah, Bena, Heidmann, Li,
  and Mayerson}}]{counterexamples-microstate-geometries}
\bibinfo{author}{\bibfnamefont{I.}~\bibnamefont{Bah}},
  \bibinfo{author}{\bibfnamefont{I.}~\bibnamefont{Bena}},
  \bibinfo{author}{\bibfnamefont{P.}~\bibnamefont{Heidmann}},
  \bibinfo{author}{\bibfnamefont{Y.}~\bibnamefont{Li}}, \bibnamefont{and}
  \bibinfo{author}{\bibfnamefont{D.~R.} \bibnamefont{Mayerson}}
  (\bibinfo{year}{2021}), \eprint{2104.10686}.

\bibitem[{\citenamefont{Bena and Mayerson}(2020)}]{Bena:2020see}
\bibinfo{author}{\bibfnamefont{I.}~\bibnamefont{Bena}} \bibnamefont{and}
  \bibinfo{author}{\bibfnamefont{D.~R.} \bibnamefont{Mayerson}},
  \bibinfo{journal}{Phys. Rev. Lett.} \textbf{\bibinfo{volume}{125}},
  \bibinfo{pages}{22} (\bibinfo{year}{2020}), \eprint{2006.10750}.

\bibitem[{\citenamefont{Bena and Mayerson}(2021)}]{Bena:2020uup}
\bibinfo{author}{\bibfnamefont{I.}~\bibnamefont{Bena}} \bibnamefont{and}
  \bibinfo{author}{\bibfnamefont{D.~R.} \bibnamefont{Mayerson}},
  \bibinfo{journal}{JHEP} \textbf{\bibinfo{volume}{03}}, \bibinfo{pages}{114}
  (\bibinfo{year}{2021}), \eprint{2007.09152}.

\end{thebibliography}
	\end{document}